\documentclass[aps,prl,twocolumn,preprintnumbers,superscriptaddress]{revtex4}
\usepackage{amssymb}
\usepackage{bm}
\usepackage{amsmath}
\usepackage[dvips]{graphicx}
\usepackage{color}

\begin{document}

\title{Diagonal charge modulation in the insulating La$_{2-x}$Sr$_x$CuO$_4$}
\author{O. P. Sushkov}
\affiliation{School of Physics, University of New South Wales, Sydney 2052, Australia}

\begin{abstract}
We show that, due to the Dzyaloshinskii-Moriya and the $XY$ anisotropies,
the disordered diagonal spin spiral in the insulating 
La$_{2-x}$Sr$_x$CuO$_4$ generates a diagonal charge density wave (CDW) with
the wave vector twice of that of the spin spiral.
The amplitude of the CDW depends on values of the
anisotropies, the doping level, and on the density of states at
the chemical potential.
Based on available experimental data  we estimate  that for 4\% doping the
amplitude of CDW is about 1/10 of the doping level.
We believe that this mechanism explains the CDW observed recently in
Zn-codoped detwinned La$_{2-x}$Sr$_x$CuO$_4$.
\end{abstract}

\date{\today}
\maketitle

The three-dimensional antiferromagnetic 
N\'{e}el order in La$_{2-x}$Sr$_{x}$CuO$_{4}$ (LSCO)
disappears at doping $x\approx 0.02$ and gives way to the so-called
spin-glass phase which extends up to $x\approx 0.055$. In both the N\'{e}el
and the spin-glass phase the system essentially behaves as an Anderson
insulator and exhibits only hopping conductivity~\cite{Keimer92,ando02}.
Superconductivity then sets in for doping $x\gtrsim 0.055$.
The incommensurate magnetic order has been observed at low temperature in 
elastic and inelastic neutron scattering. 
According to experiments in the N\'{e}el phase, the incommensurability is
almost doping-independent and directed along the orthorhombic $b$ 
axis~\cite{matsuda02}. In the spin-glass phase, the shift is also directed along the $b
$ axis, but scales linearly with doping~\cite{wakimoto99,matsuda00,fujita02}. 
Finally, in the underdoped superconducting region ($0.055\lesssim
x\lesssim 0.12$), the shift still scales linearly with doping, but it is
directed along one of the crystal axes of the tetragonal lattice~\cite{yamada98}.

In the present work we discuss only the insulating spin-glass phase, $0.02 \leq x\leq 0.055$.
The theory for the insulating phase has been suggested
in Refs.~\cite{sushkov05,luscher06,luscher07}. 
The theory has the following essential components:\newline
1)Due to strong antiferromagnetic correlations, the minima of dispersion of
a mobile hole are at points $(\pm \pi /2,\pm \pi /2)$ of the Brillouin zone,
so the system can, to some extent, be considered as a two valley semiconductor.
The hole does not have a usual spin, but it possesses a pseudospin that describes
how the hole wave function is distributed between two magnetic sublattices.
\newline
2)At low temperature, each hole is trapped in a hydrogen-like bound state
near the corresponding Sr ion, the binding energy is about $10-15\,\text{meV}$
and the radius of the bound state is about 10 \AA .\newline
3)Due to the orthorhombic distortion of LSCO, the  matrix elements 
$t_{a}^{\prime }$ and $t_{b}^{\prime }$ describing the
diagonal hopping of the hole
are slightly different, and this
makes the $b$ valley, $(-\pi /2,\pi /2)$, deeper than the $a$
valley, $(\pi /2,\pi /2)$. So all the hydrogen-like bound states are built
with holes from the $b$-valley.
In what follows, we refer to these bound states as impurities.
\newline
4)Each impurity creates a spiral distortion of the spin
background in the orthorhombic b-direction.
The distortion is observed in neutron scattering.
So the state at $0.02 < x < 0.055$ is not a spin glass, it is a
disordered spin spiral.\newline
5)At the point of overlapping of bound states (``percolation'' point) the direction of
the spiral must rotate from the diagonal to parallel because the Pauli principle.
Simultaneously the superconducting pairing is getting possible.
Hence we conclude that $x=0.055$ is the percolation point.

Intrinsically this picture does not contain any charge ordering
and this is qualitatively different from the stripe  scenario~\cite{kivelson1998}.
Charge modulation in the spin spiral picture is certainly possible,
but this can only be a secondary effect that is due to the spin-orbit
interaction.
Stimulated by the recent discovery~\cite{abbamonte2007} of the charge modulation in 
La$_{1.95}$Sr$_{0.05}$Cu$_{0.95}$Zn$_{0.05}$O$_4$ we suggest in the present
work a specific mechanism for CDW that is due to the Dzyaloshinskii-Moriya (DM)
and the $XY$ anisotropies.

{\it The spiral pitch}.
Calculation of the spiral pitch can be performed within the mean-field 
approximation.
Mobile holes are trapped  by Sr ions in hydrogen-like bound states (``impurities'') .
The ground state of the ``hydrogen atom'' is four-fold degenerate:
(two-fold pseudospin)$\times$(two-fold valley).
The orthorhombic distortion lifts the valley degeneracy, so all the impurities
reside in the b-valley~\cite{luscher06,ARPES}.
Impurity pseudospin interacts with the spiral distortion of the
spin fabric~\cite{sushkov05}.
The  interaction energy is $\sqrt{2}g Q_b$ where 
${\bf Q}$ is the wave vector of the spiral,
$Q_b$ is the component along the orthorhombic b-direction,
and  the coupling constant is approximately equal to the
antiferromagnetic exchange, $g \approx J \approx 140meV$.
We set the tetragonal lattice spacing equal to unity, so the wave vector $Q$ is dimensionless.
The pseudospin degeneracy is lifted
as soon as the spiral is established, all pseudospins
are aligned and the corresponding energy gain per unit area is 
$-x\sqrt{2}g Q$. Here $x$ is concentration
of impurities that is practically equal to doping.
The elastic energy of the spin fabric deformation is $\rho_sQ^2/2$ . 
Here $\rho_s\approx 0.18J$ is the spin stiffness. Thus the total energy is
$\rho_sQ^2/2-x\sqrt{2}g Q_b$. Minimization with respect to $Q$ gives
\begin{equation}
\label{Q}
Q=Q_b=\frac{\sqrt{2}g}{\rho_s}x \ .
\end{equation}
To fit the experimental data~\cite{wakimoto99,matsuda00,fujita02}  we need
$g=0.7J$ that agrees with the $t-J$ model estimate, $g \approx J$.
The presented mean-field picture does not address the stability of the state, broadening of the
line due to disorder, topological defects, etc. These issues have been studied in 
Refs.~\cite{luscher06,luscher07}. The stability depends on the
localization length (size of the impurity) that does not appear in the
mean-field picture. However, as soon as we know that the disordered spiral state is stable,
then the above description is correct.
We would like to stress that the spiral picture does not necessary assume
a static spiral. The spiral can be dynamic. In particular in a pure 2D system the spiral
is dynamic at any  nonzero temperature. In LSCO, due to anisotropies and
a weak 3D coupling, the spiral becomes dynamic at a small finite temperature, 
$T \sim 20K$.
However, the absence of the static spiral at $T \ge 20K$ does not mean that the spiral
is not there, it just becomes dynamic.
The important components in the above picture are 1)the hole binding, 
2)the a-valley depopulation, and 3)the height of the spin-wave dome 
$E_{cross}$ observed in neutron scattering.
Both the binding energy and the valley anisotropy energy are about 
10-15meV~\cite{sushkov05,ARPES}. The value of $E_{cross}$ depends on doping,
and  for $x=0.03 - 0.05$ it is also about 15mev~\cite{yamada2007}.
Therefore the spiral description is valid  up to characteristic temperature $T_h\sim 150K$.

\begin{figure}[h]
\centering
\par
\includegraphics[height=140pt, keepaspectratio=true]{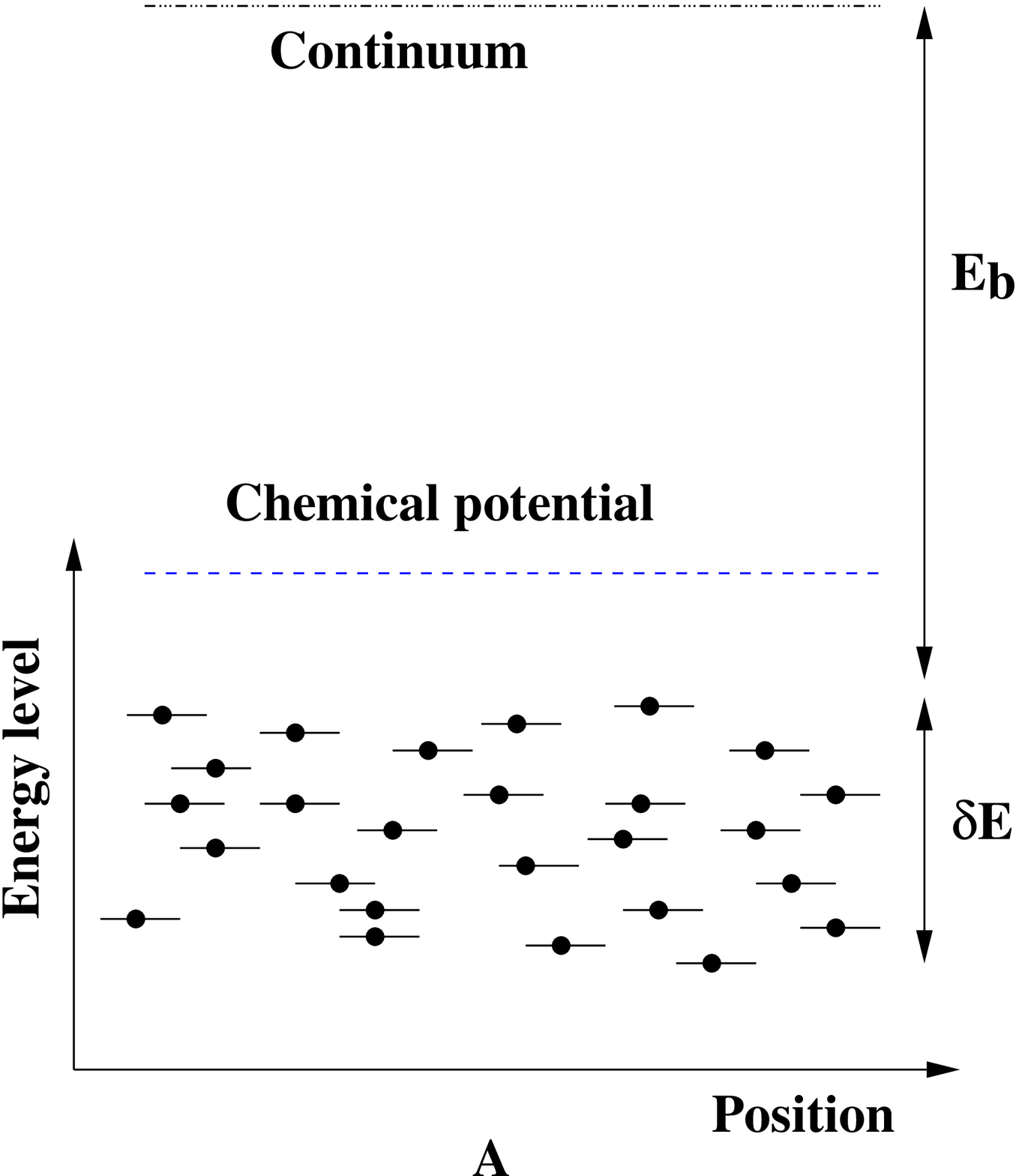}
\includegraphics[height=140pt, keepaspectratio=true]{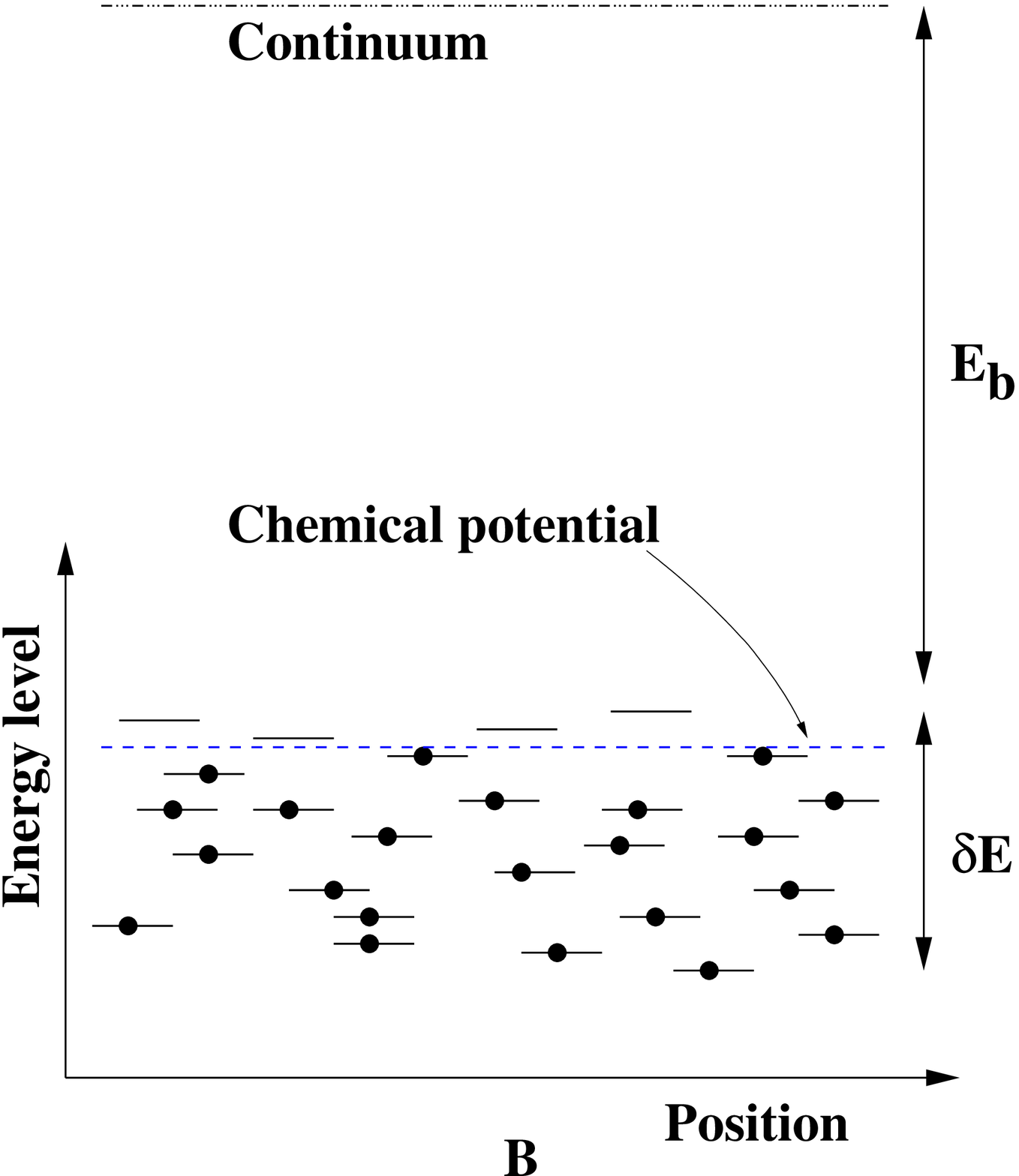}

\par
\caption{The diagram of impurity energy levels.
{\bf A}: Chemical potential is outside of the impurity band.
{\bf B}: Chemical potential is inside the impurity band.
$E_b\sim 10-15meV$ is the binding energy, and $\delta E$ is the width of
the impurity band.
}
\label{bands}
\end{figure}

{\it Density of states}.
In the case of uniform doping the diagonal spiral (unlike the parallel spiral)
always has a tendency towards charge modulation~\cite{CM,sushkov04}.
In the case of the disordered state, the problem of charge instability was resolved
in Refs.~\cite{sushkov05,luscher06,luscher07}, assuming that at zero temperature
all the Sr-hole bound states are filled and hence there is no room for compression.
So, implicitly the picture of energy levels shown in Fig.\ref{bands}A was assumed:
all the bound states are below chemical potential.
However, this picture would imply the activation behavior of dc conductivity
$\sigma \propto  \exp\{-(\Delta E/T)\}$ while it is well established
that the conductivity follows the 2D version of the  Mott  variable range hopping  
(VRH) formula~\cite{Keimer92}
\begin{equation}
\label{vrh}
\sigma \sim \exp\{-(T_0/T)^{1/3}\}\ .
\end{equation}
This implies that the chemical potential is within the range of impurity energies,
as it is shown in Fig.\ref{bands}B. Hence some bound states are unoccupied and this
gives room for CDW built on the bound states.
Let us denote the concentration of unoccupied bound states 
by $\delta x$.
It is well established that the hole doping level is pretty close to concentration of 
Sr ions, therefore, $\delta x \ll x$. On the other hand $\delta x$ is not that small
because it is sufficient for VRH. It is reasonable to assume that
\begin{equation}
\label{dx}
\frac{\delta x}{x} \sim 0.1 \ . 
\end{equation}
This is the maximum possible relative amplitude for charge density modulation,
there are no more quantum states within the impurity band to develop a larger amplitude.

The characteristic VRH temperature $T_0$ in Eq.~(\ref{vrh}) depends on doping and 
sample quality and
generally decreases when doping increases (and thus conduction becomes easier).  
At 4\% doping the data of Ref.~\cite{Keimer92} are well
fit with $T_0 \approx 500 \ \mbox{K}$ \cite{Lai}. 
Analyzing the curves from  Ref.~\cite{ando02} we  have found
$T_0(x=0.02)\sim 8000K$, $T_0(x=0.03)\sim 2000K$, $T_0(x=0.04) \sim 300K$.
The temperature $T_0$ is related to the 2D density of states $G$, 
\begin{equation}
\label{G}
G=\frac{13.8}{T_0 l^2} \,  ,
\end{equation}
see Ref.~\cite{Efros}.
Here $l$ is the localization length (the ``Bohr radius'' of the bound state).
We set the Boltzmann constant equal to unity.
The localization length is about 2.5 lattice spacings~\cite{Keimer92}.
Note that the density of states determined in this way is valid up to
$T \sim E_b \sim 100-150K$ (ionization of bound states) in spite of the fact that 
the VRH resistivity formula is valid only up to  $T \sim 20-30K$.
The density of states is related to the width of the impurity band $\delta E$
shown in Fig.\ref{bands}, 
$G=x/\delta E$.
From here we find values of $\delta E$ at different doping levels:
$\delta E(x=0.02)\sim 6meV$,
$\delta E(x=0.03)\sim 2meV$,
$\delta E(x=0.04)\sim 1meV$.

{\it Spin anisotropies and generation of the second harmonics of the spiral}.
It is very convenient to use the $\sigma$-model
notation. The energy density  in this notation reads~\cite{luscher07}
\begin{eqnarray} 
\label{eq:en}
&&\frac{\rho_s}{2} \left({\bm \nabla}{\vec n}\left({\bf r}\right)\right)^2 
- \sqrt{2}g \sum_i
{\vec \xi}_i\cdot\left[{\vec n} \times ({\bf e}_{b}\cdot{\bm \nabla}){\vec n}\right]
\delta({\bm r}-{\bm r}_i) \nonumber \\
&&+ \frac{\rho_s}{2c^2}\left[D^2 n_a^2\left({\bf r}\right)+
\Gamma_c n_c^2\left({\bf r}\right)\right] \ .
\end{eqnarray}
Here ${\vec n}$, $n^2=1$, is the staggered field that describes spins,
${\vec \xi}_i$ is direction of pseudospin of i-th impurity, $\xi^2=1$,
${\bm e}_b$ is the unit vector along the orthorhombic b-axis.
As we have already mentioned, we neglect here the impurity size.
We mention once again that what we call ``the impurity'' is the occupied bound state.
The last two terms in (\ref{eq:en}) describe anisotropies induced by the
spin-orbit interaction~\cite{chovan00,silvaneto06}.
The anisotropies ``want'' to direct ${\vec n}$ along the b-axis.
The DM vector is $D\approx 2.5meV$ and the XY anisotropy
$\sqrt{\Gamma_c} \approx 5 meV$. The spin wave velocity is $c\approx \sqrt{2}J$.

As we described above, the two first terms in (\ref{eq:en}) generate the
spiral 
\begin{equation}
\label{n}
{\vec n}=(0,\sin({\bm Q}\cdot{\bm r}+\varphi),\cos({\bm Q}\cdot{\bm r}+\varphi))\ .
\end{equation}
Here we assume that the spins are in the bc-plane.
However, we will argue below that the actual plane of the spiral is not 
important, moreover, spins can slowly fluctuate without any static spiral.
The disorder and topological defects give random phase $\varphi$ that broadens the 
line~\cite{luscher07}, however the broadening is a separate issue and we disregard it 
here. 
The XY-term in (\ref{eq:en}) is 
$\propto n_c^2 =\left[\cos({\bm Q}\cdot{\bm r}+\varphi)\right]^2\to
\frac{1}{2}\left[1+\cos(2{\bm Q}\cdot{\bm r}+2\varphi)\right]
\to-\varphi\sin(2{\bm Q}\cdot{\bm r})$.
Here we have assumed that $\varphi \ll 1$.
Thus, the interaction does not vanish after integration over space only if
\begin{equation}
\varphi =A \sin(2{\bm Q}\cdot{\bm r}) \ .
\end{equation}
This is the mechanism for generation of the second harmonics in the spin pattern.
Substitution of (\ref{n}) in (\ref{eq:en}) gives the following energy density
\begin{eqnarray}
\label{en:eq1}
\frac{\rho_s}{2}\left(Q+\varphi'\right)^2-\sqrt{2}g\sigma(Q+\varphi')
-\frac{\rho_s\Gamma_c}{2c^2}\varphi\sin(2{\bm Q}\cdot{\bm r}) \ ,
\end{eqnarray}
where $\varphi'=({\bm e}_b\cdot{\bm \nabla})\varphi$ and 
$\sigma({\bm r})=\sum_i\delta({\bm r}-{\bm r}_i)$
is the density of impurities per unit area of the plane.
In the case of distribution of states shown in Fig.~\ref{bands}A
the density of impurities is constant because there is no room for a
density modulation, $\sigma=x$.
In this case minimization of (\ref{en:eq1}) with respect to the amplitude
of the second harmonics of the spin spiral gives
\begin{equation}
\label{A}
 A=\frac{\Gamma_c}{8c^2Q^2} \ .
\end{equation}
The value of $A$ is small: $A\sim 10^{-3} - 10^{-2}$, so it is hardly
possible to observe it directly in neutron scattering. 

{\it Charge density wave}.
We know that the correct picture of states is shown in Fig.~\ref{bands}B
and in this case the CDW is possible. According to Eq.~(\ref{eq:en}) the energy 
of a single impurity is shifted due to the second harmonics of $\varphi$ as
\begin{equation}
\label{de1}
\delta\epsilon=-\sqrt{2}g\varphi'=-2\sqrt{2}g Q A\cos(2{\bm Q}\cdot{\bm r})\ .
\end{equation}
Hence, the variation of density of impurities  is
\begin{eqnarray}
\label{B}
\delta \sigma&=&-G\delta\epsilon=B\cos(2{\bm Q}\cdot{\bm r}) \ ,\nonumber\\
B&=&2G\sqrt{2}g Q A =\frac{13.8}{8 l^2 x}\frac{\Gamma_c\rho_s}{J^2T_0}
\approx \frac{0.3}{l^2x}\frac{\Gamma_c}{JT_0} \ .
\end{eqnarray}
Here $G$ is the density of states given by Eq.~(\ref{G}).
Substituting numerical values of parameters in (\ref{B}) we find that at $x=0.04$
the amplitude of charge modulation is 
\begin{equation}
\label{BB}
B\sim  10^{-2} \ .
\end{equation}
This value is in units of elementary charge per unit cell of the
square lattice.
Thus, the modulation is about 20\% of the doping level.
This dramatic enhancement of CDW is due to the very high density of states (\ref{G}).
In doing the estimate we assumed the static spin spiral in the bc-plane, see Eq.~(\ref{n}).
However, since the effect comes from the phase $\varphi$, the static order
is not essential.
In the case of fully dynamic spiral we have to replace in (\ref{B})
$\Gamma_c \to \frac{1}{2}\left(\Gamma_c+D^2\right)$. 
Since $\Gamma_c \sim D^2$, this does not influence the estimate (\ref{BB}).
According to (\ref{BB}) the CDW is so strong that it is quite possible that the real 
limitation on the amplitude $B$ comes from the available Hilbert space limit 
given by Eq.~(\ref{dx}).
Anyway, both estimates (\ref{BB}) and (\ref{dx}) give CDW amplitude of about
10\% of the doping level.

Interestingly, a naive selfconsistent treatment of the mean-field Eq.~(\ref{en:eq1})
gives an instability with respect to unlimited increasing of the CDW amplitude.
However we know from the analysis~\cite{luscher06,luscher07} that the stability issue
cannot be resolved within the mean-field approximation. This is why here we rely
on perturbation theory.
In any case, the amplitude is bound from the top by the condition (\ref{dx}) and since
the perturbation theory result (\ref{BB}) is of the same value as the upper bound we
believe that this is a reliable estimate of the effect.
The effect depends on temperature  due to depopulation
of the b-valley as well as due to ionization of bound states.
The expected dependence is roughly $B\propto \tanh(T_h/2T)$.
The characteristic temperature is $T_h \sim 150K$.

{\it Coulomb interaction, phonons,  and large correlation lengths}.
The mechanism considered above explains the CDW amplitude, but it does not
explain large correlations lengths observed in~\cite{abbamonte2007}.
One needs an additional weak interaction to coordinate the phase of the CDW.
There are two candidates for the ``coordination interaction'',
1)Coulomb interaction, 2)Interaction with phonons (lattice deformation).
Let us first consider the Coulomb interaction.
The dielectric constant $\kappa$ is strongly anisotropic~\cite{chen}.
For the direction along the c-axis it is equal to the ionic value 
$\kappa_c\sim 30-70$.
The in-plane value is much larger because polarizabilities of impurities
contribute to the screening of the electric field~\cite{chen}. This contribution
is proportional to doping, and extrapolating from data~\cite{chen} we get that
at $x=0.04$ the value is $\kappa_{ab}\sim 2000$.
Therefore, for estimates we will use the effective isotropic dielectric constant
that is average between $\kappa_c$ and $\kappa_{ab}$,
$\kappa \sim 1000$. Here we have in mind the zero temperature value.

If the Coulomb interaction is important then it must establish the CDW antiphase between the
CuO$_2$ layers. On the other hand we know that the in-plane modulation is 
very slow, $2\pi/(2Q) \ll d_c$, where  $d_c=13.15\AA$ is the separation between the planes.
A straightforward electrostatic calculation  shows that
the Coulomb energy per unit area of the plane in this geometry is 
$E_C\approx\frac{1}{2\kappa}3.2 d_c (e\delta\sigma)^2$ \ ,
 where $e\delta\sigma$ is the 
charge density per unit area, $e$ is the elementary charge and $\kappa$ is the
dielectric constant (we use CGS units).
The density wave is $\delta\sigma=B \cos(2{\bm Q}\cdot{\bm r})$.
Therefore, with account of the electrostatic energy the impurity energy shift is changed from 
(\ref{de1}) to
\begin{equation}
\label{de2}
\delta\epsilon=-2\sqrt{2}g Q A\cos(2{\bm Q}\cdot{\bm r})
+3.2\frac{e^2}{\kappa}d_c B \cos(2{\bm Q}\cdot{\bm r})\ .
\end{equation}
The density variation is $\delta\sigma=-G\delta\epsilon$. Hence, we find from 
(\ref{de2}) that the density wave amplitude is
\begin{eqnarray}
\label{B1}
B=\frac{2G\sqrt{2}g Q A}{1+3.2(e^2/\kappa)d_c G} \ .
\end{eqnarray}
The difference from Eq. (\ref{B}) is in the Coulomb factor $F_C=[1+3.2(e^2/\kappa)d_c G]^{-1}$.
The value of this factor at $x=0.04$ is $F_C\sim 0.5$.
Thus, the Coulomb interaction does not qualitatively influence the estimate 
(\ref{BB}). It can change the estimate by at most a factor  $\sim 2$.
This conclusion is supported by
the following experimental observation: The CDW amplitude is not very sensitive to
temperature up to $T\sim 100-150K$~\cite{abbamonte2007}.
On the other hand, the screening must be very sensitive to temperature
because at $T=0$ the system is an Anderson insulator while at $T \geq 50-70K$
it behaves like a conductor~\cite{ando02}. 
Thus, the CDW amplitude is not sensitive to the change of the screening regime
and hence the Coulomb interaction is not important.
The most powerful confirmation of this point comes directly from 
experiment~\cite{abbamonte2007}. In the observed CDW the layers are in-phase
and this implies that the Coulomb interaction is negligible.

Thus, as it has been pointed out in Ref~\cite{abbamonte2007}, we are left with phonons to 
coordinate the CDW phase . Most likely this is also related to the DM interaction.
The DM vector $D$ is proportional to the oxygen octachedra tilting angle,
so we can write $D\to D+\delta D$, where $\delta D$ is coupled to the
soft phonon responsible for variation of the tilting angle.
The $D^2n_a^2$-term in (\ref{eq:en}) generates coupling to the spiral
$D^2n_a^2\to(D+\delta D)^2[\sin({\bm Q}\cdot{\bm r}+\varphi)]^2 \to
D\delta D \cos(2{\bm Q}\cdot{\bm r}+2\varphi)$. This generates the lattice deformation
at the second harmonics and this is the ``coordination'' interaction.
Clearly, this mechanism gives the same phase of the CDW for nearest CuO$_2$ 
layers.

{\it In conclusion},
the disordered diagonal spin spiral in the insulating phase of LSCO
generates a charge density wave with the wavelength half 
 that of the spin spiral.
The charge modulation is due to relativistic Dzyaloshinskii-Moriya
and the $XY$ anisotropies. 
At $x=0.04$ we estimate the amplitude  of the modulation at $\sim 10\%$ of the doping level.
The effect survives up to characteristic temperature $T_h\sim 150K$.
We believe that this theory explains the CDW observed in Ref.~\cite{abbamonte2007}.

I am grateful to P.~Abbamonte and K.~Yamada
for communicating their results prior to publication.
I am also grateful to V.~Kotov and A. I. Milstein for helpful discussions.

\end{document}